\documentclass[aps,prc,twocolumn,showpacs,superscriptaddress]{revtex4}
\usepackage[latin1]{inputenc}
\usepackage{bm}
\usepackage{epsfig}
\usepackage{amsthm}
\usepackage{amsfonts}
\usepackage{float}
\usepackage{amsmath,amssymb}
\usepackage{color}
\usepackage{graphicx}% Include figure files
\usepackage{dcolumn}% Align table columns on decimal point
\usepackage{bm}% bold math
\def\nn{\nonumber}
\newcommand{\be}{\begin{equation}}
\newcommand{\ee}{\end{equation}}
\newcommand{\bea}{\begin{eqnarray}}
\newcommand{\eea}{\end{eqnarray}}

\newcommand{\ep}{\epsilon}

\newcommand{\om}{\omega}

\newcommand{\ov}{\overline}

\newcommand{\vk}{\vec k}

\newcommand{\vl}{\vec l}

\newcommand{\bk}{\boldsymbol{k}}

\newcommand{\del}{\partial}

\newcommand{\unit}{1\!\!1}
\begin{document}
\title{Bulk viscosity for pion and nucleon thermal fluctuation\\ in the hadron resonance gas model}
\author{Sabyasachi Ghosh}
\email{sabyaphy@gmail.com}
\affiliation{Department of Physics, University of Calcutta, 92, A. P. C. R
oad, Kolkata - 700009, India}
\affiliation{School of Physical Sciences, National Institute of Science Education and 
Research, Jatni, 752050, India}
\author{Sandeep Chatterjee}
\email{sandeepc@niser.ac.in}
\affiliation{School of Physical Sciences, National Institute of Science Education and 
Research, Jatni, 752050, India}
\affiliation{Theoretical Physics Division, Variable Energy Cyclotron Centre,
1/AF, Bidhan Nagar, Kolkata - 700064, India}
%
%\author{Victor Roy}
%\email{victor.physics.pm@gmail.com}
%\affiliation{Institut f\"ur Theoretische Physik, Johann Wolfgang Goethe-Universit\"at, 
%Max-von-Laue-Str. 1, D-60438 Frankfurt am Main, Germany}
%
\author{Bedangdas Mohanty}
\email{bedanga@niser.ac.in}
\affiliation{School of Physical Sciences, National Institute of Science Education and 
Research, Jatni, 752050, India}
\begin{abstract}
We have calculated microscopically bulk viscosity of hadronic matter, where
equilibrium thermodynamics for all hadrons in medium are described by Hadron
Resonance Gas (HRG) model. Considering pions and nucleons as abundant medium
constituents, we have calculated their thermal widths, which inversely control
the strength of bulk viscosities for respective components and represent their 
in-medium scattering probabilities with other mesonic and baryonic resonances, 
present in the medium. Our calculations show that bulk viscosity increases with 
both temperature and baryon chemical potential, 
%whereas when divided by the entropy density, the ratio increases first and then decreases.
whereas viscosity to entropy density ratio decreases with temperature
and with baryon chemical potential, the ratio increases first and then decreases.
The decreasing nature of the ratio with temperature is observed in most of the earlier 
investigations with few exceptions. We find that the temperature dependence 
of bulk viscosity crucially depends on the structure of the relaxation time. 
Along the chemical freeze-out line in nucleus-nucleus collisions with increasing 
collision energy, bulk viscosity as well as the bulk viscosity to entropy density 
ratio decreases, which also agrees with earlier references. Our results indicate 
the picture of a strongly coupled hadronic medium.
\end{abstract}
\maketitle
%
%
% 
%%%%%%%%%%%%%%%%%%%%%%%%%%%%%%%%%%%%%%%%%%%%%%%%%%%%%%%%%%%%%%%%%%%%%%%%%%%
\section{Introduction}
\label{sec:intro}

The extraction of the transport properties of the strongly interacting medium 
created in heavy ion collision (HIC) experiments is currently a very active topic 
of research in the HIC community. The methods of relativistic hydrodynamics with 
minimal viscous correction have been quite successful in describing the time evolution 
of the hot and dense fireball created in the HIC experiments. These kind of investigations have 
also concluded that the shear viscosity ($\eta$) to entropy 
density ($s$) ratio, $\eta/s$, of the medium created in HIC experiments is 
very close to its quantum lower bound $1/4\pi$~\cite{KSS}. 
Similar to $\eta$, another transport coefficient is the bulk
viscosity, $\zeta$, which is defined as the proportionality constant between the 
non-zero trace of the viscous stress tensor to the divergence of the fluid velocity,
and usually it appears associated with processes accompanied by a change in fluid 
volume or density. The viscous coefficient $\zeta$ has received much less attention 
than the $\eta$ in hydrodynamical simulations because its numerical value 
is assumed to be very small, as it is directly proportional to the trace of the 
energy-momentum tensor, which generally vanishes for conformally symmetric matter~\cite{Tuchin}.
However, according to Lattice Quantum Chromo Dynamics (LQCD) calculations~\cite{Lat1},
the trace of the energy momentum tensor of hot QCD medium might be large near 
the QCD phase transition, which indicates the possibility of a non-zero and large
value of $\zeta$ as well as of $\zeta/s$ near the transition temperature. This indication
is confirmed by the Refs.~\cite{LQCD_zeta1,LQCD_zeta2}, related with LQCD estimation,
where Ref.~\cite{LQCD_zeta2} exposes the possibility of divergence of $\zeta$ near 
the transition temperature. In recent times, different phenomenological 
investigations~\cite{Torrieri,Monnai,Kodama,Rajagopal,Bozek,Heinz,HM,Dusling,Victor,Grassi1,Grassi2,Habich,Gale} 
demonstrated that bulk viscosity can have a non-negligible effect on heavy ion observables,
where the values of $\zeta/s$ in Ref.~\cite{Gale} is assumed to be quite large.
%
%{\bf Sandeep} : Recently, it was realized that $\zeta$ could be large near the phase transition 
%region where lattice QCD computations reveal a peak like structure in the 
%conformal symmetry breaking measure, $\Delta = \epsilon-3P$ where $\epsilon$ is the 
%energy density and $P$ is the pressure~\cite{}. Phenomenologically, $\zeta$ has been 
%found to significantly affect the hadron spectra and the elliptic flow. This motivates 
%to investigate the magnitude of $\zeta$ of the QCD medium. 

On the basis of phenomenological importance, microscopic calculations
of $\zeta$ for quark gluon plasma (QGP) and hadronic matter is a matter of contemporary
interest in the community of HIC.
A list of references are~\cite{{Arnold},{Sasaki},{Cassing},{Defu},{G_IFT},{Kinkar},{Kadam3},
{Pratt},{Dobado},{Purnendu},{Tuchin},{Vinod},{Santosh},{Gavin},
{Sarkar},{Kadam1},{Kadam2},{Sarwar},{Noronha},{Nicola}},
where Ref.~\cite{Arnold} addressed high temperature perturbative QCD calculations of $\zeta$,
Refs.~\cite{Sasaki,Cassing,Defu,G_IFT,Kinkar,Kadam3} have gone through Nambu-Jona-Lasinio (NJL) model
calculations of $\zeta$ and Refs.~\cite{Pratt,Dobado,Purnendu} provided the discussions on
Linear Sigma Model (LSM) estimation of $\zeta$. These effective QCD model calculations
~\cite{Sasaki,Cassing,Defu,G_IFT,Kinkar,Kadam3,Pratt,Dobado,Purnendu} cover both QGP and 
hadronic phases while hadronic-model calculations of 
Refs.~\cite{Sarkar,Kadam1,Kadam2,Sarwar,Noronha,Nicola}
are restricted within hadronic phase only. The present work is also addressing
the estimation of $\zeta$ in the hadronic phase only.
At vanishing baryonic chemical potential, most of the microscopic
calculations predict that $\zeta(T)$ increases but $\zeta/s(T)$ decreases
in the hadronic temperature domain. However, few exceptions are there depending
on different scenario. For example, Ref.~\cite{Purnendu} showed that 
the decreasing function of $\zeta/s(T)$ is transformed to an 
increasing function in the hadronic temperature domain, 
when its medium constituents sigma meson becomes heavier.
Similar kind of fact is also observed in Ref.~\cite{Dobado} depending
on the different nature of phase transition as well as methodological
differences of LSM calculations. In the hadronic temperature domain,
a decreasing nature of $\zeta(T)$ is observed in Ref.~\cite{Sasaki}
while Ref.~\cite{Kadam1} estimated increasing $\zeta/s(T)$. These knowledge 
from the earlier investigations suggest that the nature of $\zeta(T)$
and $\zeta/s(T)$ are still not very settled issues. Again, the numerical strength
of $\zeta$ and $\zeta/s$ from different model calculations exhibit a large
band - $\zeta\sim 10^{-5}$ GeV$^3$~\cite{Sarkar} to $10^{-2}$ GeV$^3$~\cite{Sasaki}
or, $\zeta/s\sim 10^{-3}$~\cite{Sarkar} to $10^{0}$~\cite{Sasaki}.
These uncertainty in nature as well as numerical values of $\zeta(T)$ from the earlier
investigations demand for further research on these kind of microscopic calculations.
Owing to that motivation, we have gone through a microscopic calculations of $\zeta$
and $\zeta/s$, where equilibrium situations of hadronic matter are controlled by
the standard HRG model and non-equilibrium picture of medium
constituents is introduced via quantum fluctuation of pion and nucleon in medium.
With respect to the earlier HRG calculations of $\zeta$~\cite{Kadam1,Kadam2,Sarwar,Noronha},
the main distinguishable contribution is in the non-equilibrium properties of
medium constituents, quantified by their thermal width. Assuming pions and nucleons
as most abundant constituents of medium, we have calculated their thermal width,
which are coming from their in-medium scattering with different possible mesonic
and baryonic resonances. The main formalism for this thermal width calculations
of pion and nucleon are explicitly described in the Section~\ref{sec:form}, which
is started with a brief description HRG model, handling the equilibrium part.
Next, the numerical
results are discussed in Section~\ref{sec:num} and lastly, our investigations have
been summarized and concluded in Section~\ref{sec:concl}.

\section{Formalism}
\label{sec:form}
The HRG system is an ideal gas of hadrons and their resonances 
are taken from the Particle Data Book~\cite{pdg}.
Here we consider all resonances up
to 2 GeV masses.
The recent LQCD data at zero baryon 
chemical potential $(\mu_B)$ show 
that for temperatures up to the crossover region ($150 - 160$ MeV), HRG provides a 
reasonably good description of the LQCD thermodynamics~\cite{LQCDHRG1, 
LQCDHRG2, LQCDHRG3}. All thermodynamic quantities of the HRG can be computed from the 
logarithm of total partition function 
%$Z_{HRG}\left( T,\mu_B, \mu_Q, \mu_S\right)$
\be
\ln Z_{HRG}\left( T,\mu_B,\mu_Q,\mu_S\right)=\sum_i 
\ln Z^i_s\left( T,\mu_B, \mu_Q, \mu_S\right)~,
\label{eq.ZHRG}
\ee
where 
%$Z^i_s$ 
\be
\ln Z^i_s=\frac{g_i}{2\pi^2}VT^3\sum_{n=1}^\infty 
\frac{(\mp1)^{(n+1)}}{n^4}\left( \frac{n m_i}{T}\right)^2 K_2
\left( \frac{n m_i}{T} \right)e^{n\beta \mu_i}
\label{eq.Zs}
\ee
is the single particle partition function of the $i$th hadron.
In Eq.~(\ref{eq.Zs}), $g_i$ is the degeneracy factor of ith particle
with mass $m_i$, $V$ is volume of the medium, and $K_2(..)$ is the modified
Bessel function.
Under the condition of complete chemical equilibrium, all the hadron chemical 
potentials can be expressed in terms of only three chemical potentials 
corresponding to the QCD conserved charges
\be
\mu_i=B_i{\mu_B}_i + Q_i{\mu_Q}_i + S_i{\mu_S}_i
\label{eq.mui}
\ee
where $B_i$, $Q_i$ and $S_i$ are the baryon number, electric charge and strangeness of 
the $i$th hadron. It is straightforward to compute other thermodynamic 
quantities from $Z_{HRG}$, such as pressure ($P$), energy density ($\epsilon$),
entropy density ($s$):
\bea
P &=& -\frac{T}{V}\ln Z_{HRG}\label{eq.PHRG}~,\\
\epsilon &=& \frac{1}{V}\left\{T^2\frac{\partial\ln Z_{HRG}}{\partial T} + 
\sum_i \mu_i T\frac{\partial \ln Z_{HRG}}{\partial \mu_i}\right\}\label{eq.EHRG}~,\\
s &=& \frac{1}{T}\left\{\epsilon+P-\frac{1}{V}\sum_i \mu_iT\frac{\partial \ln Z_{HRG}}
{\partial \mu_i} \right\}\label{eq.sHRG}~.
\eea
Square of the speed of sound is defined as
\be
c_s^2 = \left(\frac{\del P}{\del \epsilon}\right)_{\rho_B}\label{cs2}~,
\ee
where $\rho_B$ is net baryon density.

%We will now discuss in detail our scheme to estimate 
%the effect of the medium on the $\zeta$ of HRG. 
From the Relaxation Time Approximation (RTA) of kinetic theory approach~\cite{Purnendu,Gavin} or from 
the one-loop expression of diagrammatic approach based on Kubo formula~\cite{Nicola},
we can get standard expressions of bulk viscosity coefficient for pion and nucleon 
components~\cite{Purnendu,Gavin,Nicola,Kadam2} :
\be
\zeta_\pi = \left(\frac{g_\pi}{T}\right) \int \frac{d^3\bk}{(2\pi)^3} 
\, \frac{ n_\pi \left[1 + n_\pi\right]}{\om_\pi^2 \, \Gamma_\pi} 
\left\{\left(\frac{1}{3} - c_s^2\right) \bk^2 
- c_s^2 m_\pi^2  \right\}^2
\label{zeta_pi}
\ee
and
\bea
\zeta_N &=& \left(\frac{g_N}{T} \right) \int \frac{d^3\bk}{(2\pi)^3} 
\, \frac{1}{{\om_N^2 \, \Gamma_N}}\left[ 
\left\{ \left(\frac{1}{3} - c_s^2\right) \bk^2 - c_s^2 m_N^2 
\right.\right.\nn \\ 
&&\left.\left. 
- \om_N\left(\frac{\del P}{\del \rho_B}\right)_{\ep} \right\}^2
n^+_N \left(1 - n^+_N\right) + \left\{ \left(\frac{1}{3} - c_s^2\right) \bk^2
\right.\right.\nn\\
&& \left.\left.
 - c_s^2 m_N^2 
+ \om_N\left(\frac{\del P}{\del \rho_B}\right)_{\ep} \right\}^2
n^-_N \left(1 - n^-_N\right)\right] ~,
\label{zeta_N}
\eea

where $n_\pi=1/\{e^{\om_\pi/T}-1\}$ is the Bose-Einstein (BE) distribution
function of pion with energy $\om_\pi=\{\bk^2 + m_\pi^2\}^{1/2}$,
$n^{\pm}_N=1/\{e^{(\om_N \mp \mu_B)/T}+1\}$ are the Fermi-Dirac (FD)
distribution functions of nucleon and anti-nucleon respectively with energy 
$\om_N=\{\bk^2 + m_N^2\}^{1/2}$ at finite temperature $T$ and
baryon chemical potential $\mu_B$. The degeneracy factors
of pion and nucleon components are $g_\pi=3$ and $g_N=2\times 2$ respectively.
\begin{figure}
\begin{center}
\includegraphics[scale=0.52]{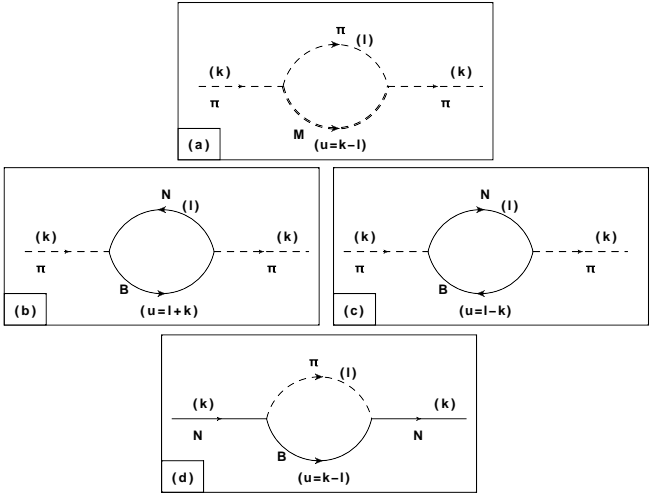}
\caption{Pion self-energy diagram with mesonic loops (a)
and baryonic loops [(b) and (c) are direct and cross diagrams] and
nucleon self-energy diagram (d).} 
\label{Bulk_pi_N}
\end{center}
\end{figure}
Next, let us come to the important quantities $\Gamma_\pi$ 
and $\Gamma_N$ of Eq.~(\ref{zeta_pi}) and (\ref{zeta_N}), 
which are called thermal widths of pion and nucleon respectively.
During propagation in the medium, pion and nucleon may go
through different on-shell scattering with other mesonic ($M$) 
and baryonic ($B$) resonances, which can be quantified by 
their different possible self-energy diagrams. 
From the imaginary part of their self-energy functions,
their respective thermal widths $\Gamma_\pi$ and $\Gamma_N$ 
can be found. Fig.~\ref{Bulk_pi_N}(a) represents pion self-energy
with internal lines of pion ($\pi$) and other mesonic resonances ($M$),
which we can shortly call $\pi M$ loop. We will take $M=\sigma$ and $\rho$,
as they are dominant resonances of $\pi\pi$ decay
channel (within the invariant mass range of 1 GeV).
Now, from the retarded self-energy of pion for $\pi M$ loop 
$\Pi^R_{\pi(\pi M)}(k)$, the corresponding thermal
width $\Gamma_{\pi(\pi M)}$ can be obtained as
\be
\Gamma_{\pi(\pi M)}=-{\rm Im}{\Pi}^R_{\pi(\pi M)}(k_0=\om_\pi,\vk)/m_\pi ~,
\label{pipiM}
\ee
where subscript notation stands for external (outside the bracket) 
and internal (inside the bracket) particles for 
the diagram~\ref{Bulk_pi_N}(a). Following Similar notation, we can
define
\be
\Gamma_{\pi(NB)}=-{\rm Im}{\Pi}^R_{\pi(NB)}(k_0=\om^\pi_k,\vk)/m_\pi ~,
\label{piNB}
\ee
where intermediate states of pion self-energy are nucleon $N$ and other
baryonic resonance $B$ as shown in Fig.~\ref{Bulk_pi_N}(b) along with
its cross diagram (c). As a dominant 4-star baryons with spin $J_B=1/2$ and $3/2$,
we have taken $B =\Delta(1232)$, $N^*(1440)$, $N^*(1520)$,
$N^*(1535)$, $\Delta^*(1600)$, $\Delta^*(1620)$, $N^*(1650)$, 
$\Delta^*(1700)$, $N^*(1700)$, $N^*(1710)$ and $N^*(1720)$.
Adding all these mesonic ($\pi M$) and baryonic ($NB$) loops, 
the total thermal width of pion $\Gamma_\pi$ can be obtained as
\be
\Gamma_\pi= \Gamma^M_\pi + \Gamma^B_\pi =
\sum_M\Gamma_{\pi(\pi M)} + \sum_B\Gamma_{\pi(NB)}~.
\label{pi_piM_NB}
\ee

Similarly, one-loop self-energy of nucleon with pion ($\pi$)
and baryon ($B$) intermediate states, which is denoted as
$\Sigma^R_{N(\pi B)}$ (retarded part), will be our matter of interest
to estimate corresponding nucleon thermal width $\Gamma_{N(\pi B)}$.
The diagramatic anatomy of $\Sigma^R_{N(\pi B)}$ is shown in 
Fig.~\ref{Bulk_pi_N}(d). Here we have taken all the 4-star 
spin $1/2$ and $3/2$ baryons, mentioned above. 
Hence, summing all the $\pi B$ loops, we can get our total nucleon thermal width : 
\be
\Gamma_N=\sum_B\Gamma_{N(\pi B)}=-\sum_B{\rm Im}\Sigma^R_{N(\pi B)}(k_0=\om_N,\vk)~.
\label{Gam_N}
\ee
The imaginary part of self-energies, given in 
Eqs~(\ref{pipiM}), (\ref{piNB}) and
(\ref{Gam_N}), have been derived with help of 
standard thermal field theoretical techniques.
At first, the expression for Im$\Pi^R_{\pi(\pi M)}$ is~\cite{GKS}
\bea
{\rm Im}\Pi^R_{\pi(\pi M)}(k_0=\om_\pi, \vk) &=& \int \frac{d^3 \vl}{32\pi^2 \om_l\om_u}
\nn\\
&&L_{\pi\pi M}(k,l)|_{(l_0=-\om_l,~k_0=\om_k)}
\nn\\
&&\left(n_l - n_u\right)~\delta(\om_\pi +\om_l - \om_u)
~,
\nn\\
\label{G_pi_piM}
\eea
where $n_l$, $n_u$ are BE distribution functions of $\pi$, $M$
mesons respectively at energies $\om_l=\{\vl^2 +m_\pi^2\}^{1/2}$
and $\om_u=\{(\vk - \vl)^2 + m_M^2\}^{1/2}$.
The vertex factors $L_{\pi(\pi M)}(k,l)$~\cite{GKS} 
%\bea
%L_{\pi(\pi M)}(k,l) &=& - \frac{g^2_\sigma m_\sigma^2}{4}, 
%~~~~~~~~~~~~~~~~~~~{\rm for}~M=\sigma~,
%\nn\\
% &=& -\frac{g^2_\rho}{m_\rho^2} \, 
%[ k^2 \left(k^2 - m^2_\rho\right) + 
%l^2 \left(l^2 - m^2_\rho\right) 
%\nn\\
%&&- \, 2\{ (k\cdot l) \, m^2_\rho + k^2 \,l^2 \}],~{\rm for}~M=\rho
%\nn\\
%\eea
have been calculated by using the effective Lagrangian density,
\be
{\cal L}_{\pi\pi M} = g_\rho \, {\vec \rho}_\mu \cdot {\vec \pi} \times \del^\mu {\vec \pi} 
+ \frac{g_\sigma}{2} m_\sigma {\vec \pi}\cdot {\vec\pi}\,\sigma~,
\label{Lag_pipiM}
\ee
where $g_\rho$ and $g_\sigma$ are respectively effective coupling constants of $\rho$
meson field $({\vec \rho}_\mu)$ and $\sigma$ meson field
($\sigma$), which are coupled with the pion field (${\vec \pi}$).  

Next, the direct and cross diagrams of pion self-energy for $NB$ loop
are combinedly expressed as~\cite{G_pi_JPG,G_eta_BJP}
\bea
{\rm Im}\Pi^R_{\pi(NB)}(k_0=\om_\pi, \vk) &=& \int \frac{d^3 \vl}{32\pi^2 \om_l\om_u}
\nn\\
&&L_{\pi NB}(k,l)|_{(l_0=-\om_l,~k_0=\om_k)}
\nn\\
&&\{\left(-n^+_l + n^+_u\right)~\delta(\om_\pi +\om_l - \om_u)
\nn\\
&&+\left(-n^-_l + n^-_u\right)~\delta(\om_\pi -\om_l + \om_u)\}
~,
\nn\\
\label{G_pi_NB}
\eea
where $n^{\pm}_l$, $n^{\pm}_u$ are FD distribution functions of $N$ and $B$
($\pm$ for particle and anti-particle)
respectively at energies $\om_l=\{\vl^2 +m_N^2\}^{1/2}$
and $\om_u=\{(\vk \pm \vl)^2 + m_B^2\}^{1/2}$ ($\pm$ for diagrams 
(b) and (c) respectively).
With the help of the effective Lagrangian densities for $\pi NB$ 
interactions~\cite{Leopold},
\bea
{\cal L}_{\pi NB}&=&\frac{f}{m_\pi}{\ov \psi}_B\gamma^\mu
\left\{
\begin{array}{c}
i\gamma^5 \\
\unit
\end{array}
\right\}
\psi_N\del_\mu\pi + {\rm h.c.}~{\rm for}~J_B^P=\frac{1}{2}^{\pm},
\nn\\
&=&\frac{f}{m_\pi}{\ov \psi}^\mu_B
\left\{
\begin{array}{c}
\unit \\
i\gamma^5
\end{array}
\right\}
\psi_N\del_\mu\pi + {\rm h.c.}~{\rm for}~J_B^P=\frac{3}{2}^{\pm}~,
\nn\\
&&(P~{\rm stands~for~parity~of~}B)
\label{Lag_BNpi}
\eea
one can deduced the vertex factors $L_{\pi NB}(k,l)$~\cite{G_pi_JPG,G_eta_BJP}.
%\bea
%L_{\pi NB}(k,l)
%&=&-4\left(\frac{f}{m_\pi}\right)^2[2(k\cdot l)^2-a(k\cdot l)k^2
%\nn\\
%&&-k^2(l^2+m_Nm_B)],~~~~~ {\rm for}~ J_B^P=\frac{1}{2}^{\pm},
%\nn\\
%&=&-\frac{8}{3m_B^2}\left(\frac{f}{m_\pi}\right)^2[m_Nm_B+l^2-a(k\cdot l)]
%\nn\\
%&&[(l\cdot k-ak^2)^2-k^2m_B^2],~ {\rm for}~ J_B^P=\frac{3}{2}^{\pm}.
%\nn\\
%\eea
%
At last, the expression for Im$\Pi^R_{N(\pi B)}$ is~\cite{G_N,G_NNst_BJP}
\bea
{\rm Im}\Pi^R_{N(\pi B)}(k_0=\om_\pi, \vk) &=& \int \frac{d^3 \vl}{32\pi^2 \om_l\om_u}
\nn\\
&&L_{N\pi B}(k,l)|_{(l_0=-\om_l,~k_0=\om_k)}
\nn\\
&&\left(n_l + n^+_u\right)~\delta(\om_\pi +\om_l - \om_u)
~,
\nn\\
\label{G_N_piB}
\eea
where $n_l$ is BE distribution functions of $\pi$ at 
energy $\om_l=\{\vl^2 +m_\pi^2\}^{1/2}$ and $n^+_u$ is
FD distribution of $B$ at energy $\om_u=\{(\vk - \vl)^2 + m_M^2\}^{1/2}$.
With the help of the interaction Lagrangian 
densities from Eq.~(\ref{Lag_BNpi}),
the vertex factors $L_{N\pi B}(k,l)$~\cite{G_N} have been obtained.
%\bea
%L_{N\pi B}(k,l)&=&-\left(\frac{f}{m_\pi}\right)^2\left\{\left(\frac{R^2}{2}-m_\pi^2
%\right)l_0
%\right.\nn\\
%&&\left.~~~-Pm_\pi^2m_B\right\}
%~{\rm for}~~~~~~~~~~J_B^P=\frac{1}{2}^{\pm}~,
%\nn\\
%&=&-\left(\frac{f}{m_\pi}\right)^2\frac{2}{3m_B^2}
%\left\{\left(\frac{R^2}{2}-m_\pi^2\right)^2
%\right.\nn\\
%&&\left.-m_\pi^2m_B^2\right\}(k_0-l_0+Pm_B)
%~~{\rm for}~J_B^P=\frac{3}{2}^{\pm}~.
%\nn\\
%\eea
% 

%%%%%%%%%%%%%%%%%%%%%%%%%%%%%%%%%%%%%%%%%%%%%%%%%%%%%%%%%%%%%%%%%%%%%%%%%%%
\section{Results and Discussion}
\label{sec:num}
\begin{figure}
\begin{center}
\includegraphics[scale=0.35]{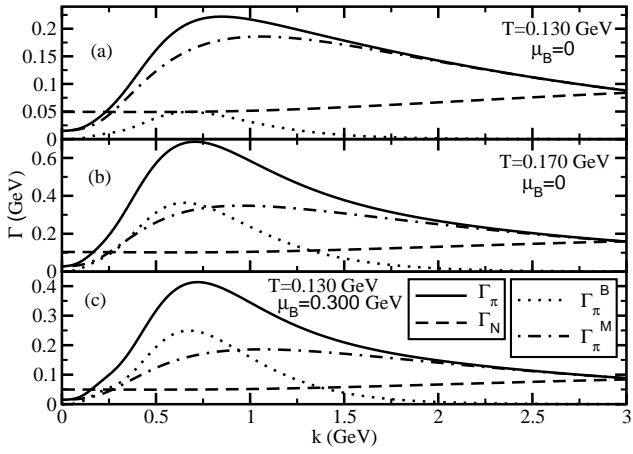}
\caption{Momentum distribution of pion thermal width for 
mesonic (dash-dotted line), baryonic loops (dotted line) 
and their total (solid line) and nucleon thermal width 
(dashed line) at three different medium parameters: 
(a) $(T,\mu_B)=(0.130$ GeV, $0)$, (b) $(0.170$ GeV, $0)$
and (c) $(0.130$ GeV, $0.300$ GeV).} 
\label{gm_k}
\end{center}
\end{figure}
%
%
%\begin{figure}
%\begin{center}
%\includegraphics[scale=0.35]{z_T_t10.eps}
%\caption{$\zeta(T)$ for $m_\pi=0$ (dotted line), $0.070$ (dashed line),
%$0.140$ GeV (solid line) at $c_S^2=0$ (a), $c_S^2=1/3$ (b)
%and $c_S^2(T)$ from HRG (c).} 
%\label{z_T_t10}
%\end{center}
%\end{figure}
%
%
%\begin{figure}
%\begin{center}
%\includegraphics[scale=0.35]{z_mpi_T.eps}
%\caption{$\zeta$ vs $m_\pi$ at different values $c_S^2$.} 
%\label{z_mpi_T}
%\end{center}
%\end{figure}
%
Let us start our numerical discussion with the Fig.~(\ref{gm_k}),
where momentum distribution of thermal widths of pion and nucleon
have been displayed. With the help of Eqs.~(\ref{pipiM}), (\ref{piNB}),
(\ref{pi_piM_NB}), (\ref{G_pi_piM}) and (\ref{G_pi_NB}), $\Gamma_\pi^M$,
$\Gamma_\pi^B$ and their total $\Gamma_\pi$ can be found whose momentum
distributions are respectively shown by dash-dotted, dotted and solid line in 
the Fig.~(\ref{gm_k}). Similarly, $\Gamma_N$ can be deduced by using 
Eqs.~(\ref{G_N_piB}) and (\ref{Gam_N}) and its momentum distribution is
represented by dash line. Panels (a), (b) and (c) of Fig.~(\ref{gm_k})
are for different set of temperature $T$ and baryon chemical potential 
$\mu_B$ of the medium. Though $\Gamma_N$ is approximately constant with 
nucleon momentum, but $\Gamma_\pi^M$ and $\Gamma_\pi^B$ exhibit a peak
structure in some point of $\vk$-axis, which depends on the medium 
parameters $T$ and $\mu_B$. These momentum distribution of $\Gamma_\pi$
and $\Gamma_N$ will be integrated out when we will estimate $\zeta_\pi$
and $\zeta_N$ from Eqs.(\ref{zeta_pi}) and (\ref{zeta_N}) respectively.

\begin{figure}
\begin{center}
\includegraphics[scale=0.35]{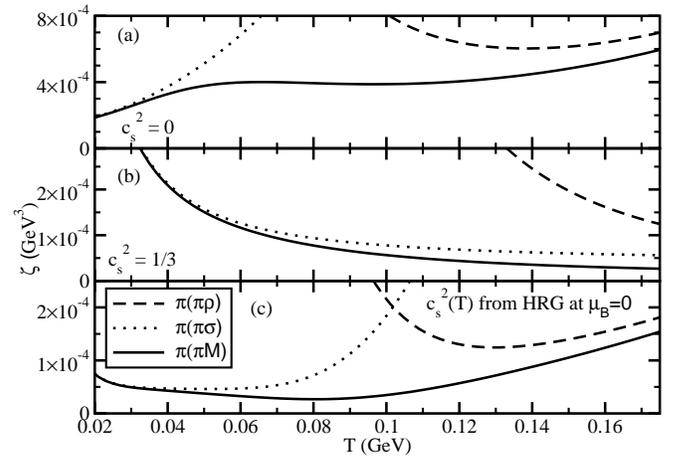}
\caption{$\zeta(T)$ due to pion thermal width for $\pi\sigma$ (dotted line),
$\pi\rho$ (dashed line) loops and their total (solid line)
at $c_S^2=0$ (a), $c_S^2=1/3$ (b) and $c_S^2(T)$ from HRG (c).} 
\label{z_T_ppM}
\end{center}
\end{figure}
\begin{figure}
\begin{center}
\includegraphics[scale=0.35]{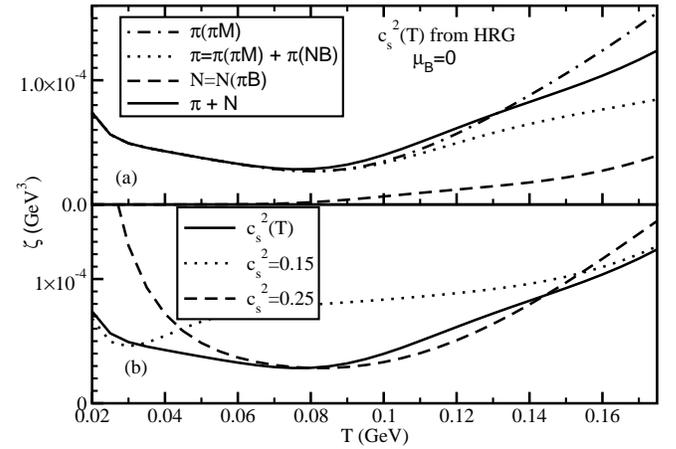}
\caption{(a): Temperature dependence of bulk viscosity for pion thermal
width with mesonic loops (dash-dotted line), meson + baryon loops 
(dotted line), for nucleon thermal width (dashed line) and their total 
$\zeta_T=\zeta_\pi +\zeta_N$ (solid line). (b): $\zeta(T)$
for $c_S^2(T)$ from HRG and two constant values of $c_S^2$ 
($c_S^2=0.15$: dotted line and $c_S^2=0.25$: dash line).} 
\label{z_T_mu0}
\end{center}
\end{figure}
%
%
%\begin{figure}
%\begin{center}
%\includegraphics[scale=0.35]{z_T_mu0cs.eps}
%\caption{(b): $c_S^2(T)$, obtained by taking full (F) set of resonance 
%(solid line), restricted (R) set of resonance (dash line) and 
%only $\pi+N$ (dotted line). LQCD results (circles), taken from \cite{LQCD_cs2}.
%(a): Corresponding bulk viscosity coefficients.} 
%\label{z_T_mu0cs}
%\end{center}
%\end{figure}
%
%
\begin{figure}
\begin{center}
\includegraphics[scale=0.35]{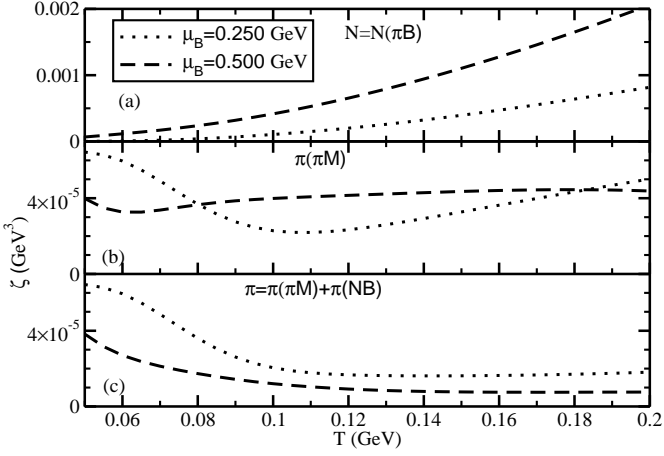}
\caption{$\zeta(T)$ due to nucleon thermal width (a), pion
thermal width for meson loops (b) and meson + baryon loops (c) at
$\mu_B=0.250$ GeV (dotted line) and $0.500$ GeV (dash line).} 
\label{z_Tmu}
\end{center}
\end{figure}
%

%Instead of constant thermal width, let us now proceed to
%use our calculated thermal width, which depends on momentum of
%medium constituents and medium parameters $T$ and $mu_N$.
Let us come to the different loop contributions of pion and
nucleon thermal width in bulk viscosity
coefficient of hadronic matter.
Fig.~\ref{z_T_ppM}(c) shows individual contributions of $\pi\sigma$
(dotted line) and $\pi\rho$ (dash line) loops in $\zeta_\pi$, which
reveals that they are respectively important in low ($T<0.080$ GeV)
and high ($T>0.080$ GeV) temperature domain for getting a non-divergent
values of $\zeta_\pi$. These are respectively obtained by putting 
$\Gamma_{\pi(\pi \sigma)}$ and $\Gamma_{\pi(\pi \rho)}$ in place of 
$\Gamma_\pi$ of Eq.~(\ref{zeta_pi}). Putting
$\Gamma_\pi^M=\Gamma_{\pi(\pi \sigma)}+\Gamma_{\pi(\pi \rho)}$  
in place of $\Gamma_\pi$ of Eq.~(\ref{zeta_pi}), we get the solid
line, representing total bulk viscosity of pionic component due
to meson loops. After a mild decrement in low $T$ ($<0.080$ GeV),
it receives an increment nature in high $T$ ($>0.080$ GeV).
Along with Fig.~\ref{z_T_ppM}(c), where an explicit temperature
dependent $c_s^2$ is taken from HRG model, the results for $c_s^2=0$
and $c_s^2=1/3$ are also displayed in Fig.~\ref{z_T_ppM}(a) and (b),
which are little different in nature. Just to show the phase space
sensitivity of bulk viscosity via $c_s^2$, these two results are displaying
two extreme limits of $c_s^2$. Therefore, we can understand 
Fig.~\ref{z_T_ppM}(c) as some sort of superposition of \ref{z_T_ppM}(a) and (b).

\begin{figure}
\begin{center}
\includegraphics[scale=0.35]{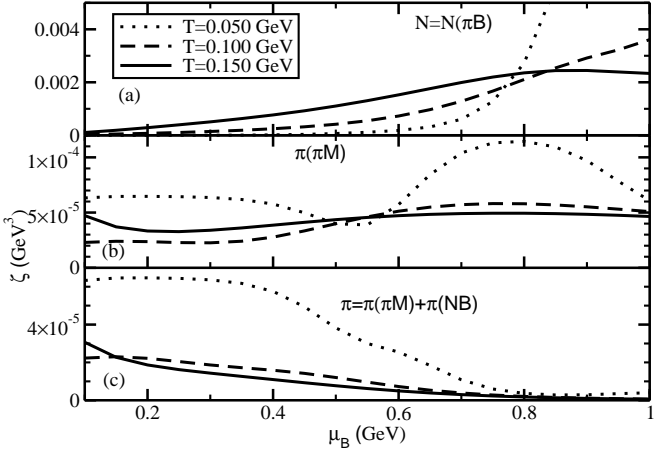}
\caption{Same as Fig.~(\ref{z_Tmu}) along $\mu_B$-axis 
at $T=0.050$ GeV (dotted line),
$0.100$ GeV (dash line) and $0.150$ GeV (solid line).} 
\label{z_mu_piN}
\end{center}
\end{figure}
\begin{figure}
\begin{center}
\includegraphics[scale=0.35]{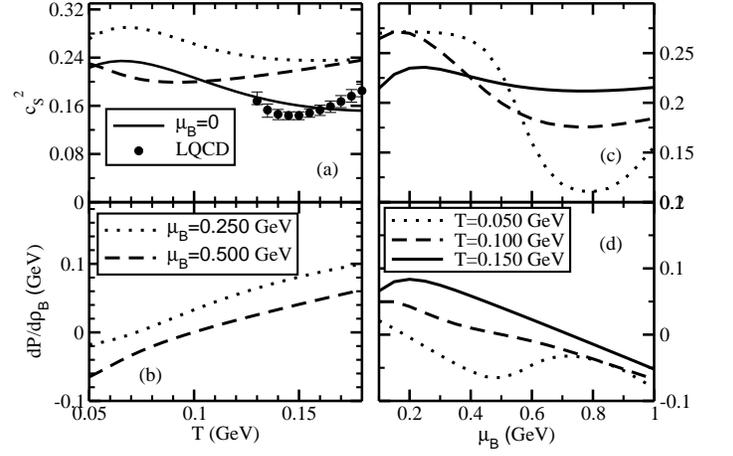}
\caption{(a): $c_S^2(T)$ at $\mu_B=0$ (solid line), $0.250$ GeV (dotted line) 
and $0.500$ GeV (dash line), and LQCD results 
of $c_s^2(T,\mu_B=0)$ (circles)~\cite{Lat1};
(b) : $\left(\frac{\del P}{\del \rho_B}\right)_{\ep}$ vs $T$ at
$\mu_B=0.250$ GeV (dotted line) and $0.500$ GeV (dash line);
(c): $c_S^2(\mu_B)$ at $T=0.050$ GeV (dotted line),
$0.100$ GeV (dash line) and $0.150$ GeV (solid line);
(d) : $\left(\frac{\del P}{\del \rho_B}\right)_{\ep}$ vs $\mu_B$ at
$T=0.050$ GeV (dotted line),
$0.100$ GeV (dash line) and $0.150$ GeV (solid line).
} 
\label{cs2_Tmu}
\end{center}
\end{figure}
\begin{figure}
\begin{center}
\includegraphics[scale=0.35]{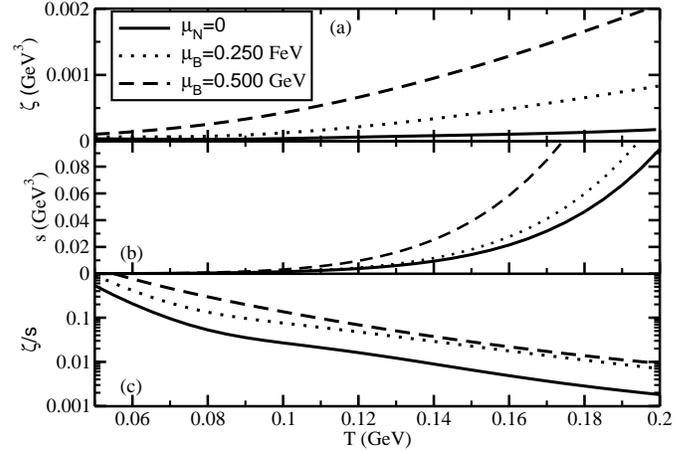}
\caption{$T$ dependence of total bulk viscosities (a), entropy densities from HRG (b) and
their ratios $\zeta/s$ (c) at $\mu_B=0$ (solid line), $0.250$ GeV 
(dotted line) and $0.500$ GeV (dash line).} 
\label{z_s_Tmu}
\end{center}
\end{figure}
\begin{figure}
\begin{center}
\includegraphics[scale=0.35]{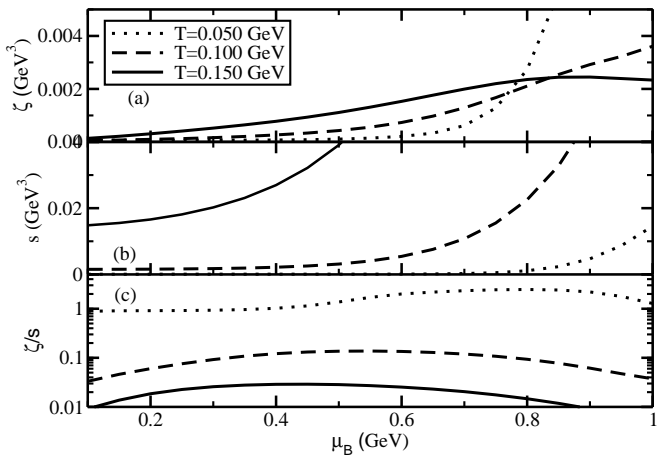}
\caption{$\mu_B$ dependence of total bulk viscosities (a), 
entropy densities from HRG (b) and
their ratios $\zeta/s$ (c) at $T=0.050$ GeV (dotted line), 
$0.100$ GeV (dash line) and $0.150$ GeV (solid line).} 
\label{z_s_mu}
\end{center}
\end{figure}
\begin{figure}
\begin{center}
\includegraphics[scale=0.35]{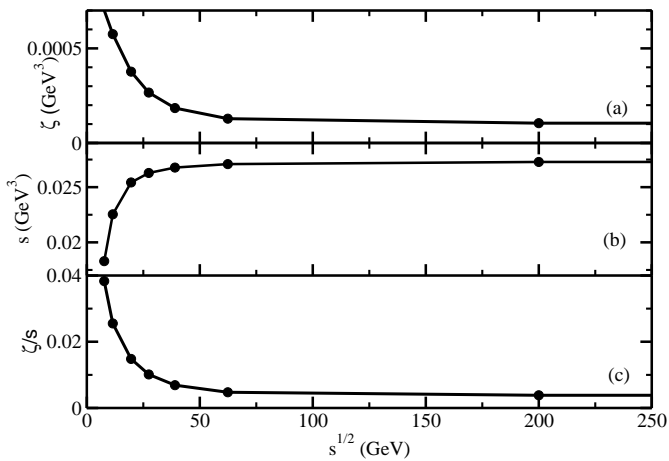}
\caption{Center of mass energy ($\sqrt{s}$) dependence of 
total bulk viscosity (a), entropy density from HRG (b) and
their ratio $\zeta/s$ (c).} 
\label{z_root_s}
\end{center}
\end{figure}
According to Eq.~(\ref{pi_piM_NB}) different baryon loops contribution 
($\Gamma_\pi^B$) should have to add with meson loops contribution 
($\Gamma_\pi^M$) to get total pion thermal width $\Gamma_\pi$. 
In Fig.~\ref{z_T_mu0}(a), changing the nature of dash-dotted line
to dotted line indicates that inclusion of baryon loops with meson
loops becomes the reason for reducing the rate of increment of 
$\zeta_\pi(T)$ at high temperature region, $T>0.100$ GeV.
Putting our calculated nucleon thermal width $\Gamma_N$ in Eq.~(\ref{zeta_N}),
we get $\zeta_N$ as shown by dash line in Fig.~\ref{z_T_mu0}(a).
Now adding $\zeta_N$ with $\zeta_\pi$ we have total bulk viscosity
\be
\zeta_T=\zeta_\pi + \zeta_N~,
\ee
as shown by solid line in Fig.~\ref{z_T_mu0}(a). In Fig.~\ref{z_T_mu0}(b),
this $\zeta_T$ (solid line) has been compared with the results generated for two
constant values of $c_s^2$ ($c_s^2=0.25$: dash line and $c_s^2=0.15$: dotted line),
within which $c_s^2(T, \mu_B=0)$ from HRG model more or less varies.

%Next in Fig.~\ref{z_T_mu0cs}(b), $c_s^2(T)$ at $\mu_B=0$,
%obtained from HRG model taking full (F) set of resonances up to 2 GeV,
%is represented by the solid line, which is well agreement with Lattice
%QCD results~\cite{LQCD_cs2} (circles) within the hadronic temperature domain 
%(0.130 GeV < T < 0.175 GeV). Restricting HRG model calculation within
%some restricted (R) resonances, which are mainly taken for loop calculations
%in pion and nucleon self-energy, we get $c_s^2(T)$ as a dash line, which are
%quite away from the LQCD values. While, thermodynamics of pion and nucleon 
%($\pi +N$) only gives $c_s^2(T)$ (dotted line), which is much far from 
%LQCD results. Therefore, full HRG model calculation for thermodynamical
%quantities like $c_s^2$ and others ($\epsilon$, $P$, $s$) are considered in
%the results of our graphs. Fig.~\ref{z_T_mu0cs}(a) shows $\zeta_T$ for three
%different values of $c_s^2(T)$ in Fig.~\ref{z_T_mu0cs}(b) and they are mainly
%little different in high temperature domain.
%
At two different values of $\mu_B$, $\zeta(T)$ due to nucleon thermal width ($\Gamma_N$), pion
thermal width for meson loops ($\Gamma_\pi^M$) and meson + baryon loops ($\Gamma_\pi$)
are shown in Fig.~\ref{z_Tmu}(a), (b) and (c) respectively. 
Similarly, Fig.~\ref{z_mu_piN}(a), (b) and (c) are displaying different loop
contributions in $\zeta(\mu_B)$ at $T=0.050$ GeV (dotted line), 
$0.100$ GeV (dashed line) and $0.150$ GeV (solid line).
From Fig.~\ref{z_Tmu}(a) and \ref{z_mu_piN}(a),
we see that $\zeta_N$ increases with $T$ as well as $\mu_B$.
%Although their decreasing nature is also noticed at low $T$ and $\mu_B$.
From Fig.~\ref{z_Tmu}(b), we see 
the $\zeta_\pi$ due to $\Gamma_\pi^M$ at finite $\mu_B$ first decreases 
at low $T$ then increases at high $T$. 
The nature of these curves are quite similar to the curve of 
$\zeta_\pi(T)$ at vanishing $\mu_B$ but their minima are only
shifted towards lower $T$ as $\mu_B$ increases. Following the 
same story of vanishing $\mu_B$, inclusion of baryon loops in 
pion self-energy is again influencing on $\zeta_\pi(T)$ in high
temperature domain. 
%It can be understood by noticing simultaneously 
%Fig.~\ref{z_Tmu}(a) for finite $\mu_B$ and \ref{z_T_mu0}(a) for $\mu_B=0$.
The variation with $\mu_B$ of $\zeta_N(\mu_B)$ in Fig.~\ref{z_mu_piN}(a)
and $\zeta_\pi(\mu_B)$ in Fig.~\ref{z_mu_piN}(b) and (c) are grossly same
as their temperature dependence. For small $T$ and $\mu_B$, $\zeta_N$ and 
$\zeta_\pi$ are of similar order. However, with increasing $T$ and $\mu_B$, 
$\zeta_N$ dominates over $\zeta_\pi$. $\zeta_N$ receives additional contribution 
from $\left(\frac{\del P}{\del \rho_B}\right)_{\ep}$.
%At $\mu_B=0$, the numerical values of $\zeta_\pi$ and $\zeta_N$ are almost 
%in same order but at finite $\mu_B$, $\zeta_N$ becomes dominant over $\zeta_\pi$. 
%This is because of the new quantity $\left(\frac{\del P}{\del \rho_B}\right)_{\ep}$, 
%which is appeared in the expression of $\zeta_N$ for non-vanishing $\mu_B$. 
One should keep in mind that the term $\left(\frac{\del P}{\del \rho_B}\right)_{\ep}$
goes to zero for $\mu_B=0$. The $T$ and $\mu_B$
dependence of $\left(\frac{\del P}{\del \rho_B}\right)_{\ep}$ are shown in 
Fig.~\ref{cs2_Tmu}(b) and (d) respectively while Fig.~\ref{cs2_Tmu}(a) and (c)
are displaying the $T$ and $\mu_B$ dependence of $c_s^2$. From Fig.~\ref{cs2_Tmu}(a),
we see that our $c_s^2(T, \mu_B=0)$ curve (solid line) is in good agreement 
with LQCD results~\cite{Lat1} (circles) 
within the hadronic temperature domain ($T < 0.160$ GeV). 
Total bulk viscosity $\zeta_T$ (a), entropy density $s$ (b) and their 
ratio $\zeta/s$ (c) are plotted against $T$ in Fig.~(\ref{z_s_Tmu}) 
and $\mu_B$ in Fig.~(\ref{z_s_mu}) 
at three different values $\mu_B$ and $T$ respectively.
Since increment of $s(T)$ is larger than the increment of $\zeta(T)$,
therefore, $\zeta/s$ is appeared as a decreasing function of $T$.
On the other hand, both $\zeta(\mu_B)$ and $s(\mu_B)$ 
monotonically increase with $\mu_B$ but 
the ratio $\zeta/s(\mu_B)$ increases first and then decreases at high 
$\mu_B$ domain. 
%This mild valley structure in $\mu_B$
%axis can also be observed for any constant values of thermal width of
%medium constituents. It means that the thermodynamical phase space,
%containing the thermal distribution functions as well as the thermodynamical
%quantities like $c_S^2(T,\mu_B)$ and 
%$\left(\frac{\del P}{\del \rho_B}\right)_{\ep}\Big(T,\mu_B\Big)$, are mainly responsible
%for this mild valley structure. Among them, impact of the 
%$\left(\frac{\del P}{\del \rho_B}\right)_{\ep}$ is largest.
Next, Fig.~\ref{z_root_s}(a), (b) and (c) reveal respectively
the variation of total bulk viscosity $\zeta$, entropy density $s$ and their
ratio with the variation of center of mass energy $\sqrt{s}$
%
%\footnote{
(Reader are
requested to be careful on the same symbol $s$ used for entropy density and square of
beam energy).
%}.
%
The beam energy 
dependence of $T$ and $\mu_B$ used in computation are those obtained from fits 
to hadron yields. We have used the parameterization from Ref.~\cite{HRGKarsch}.
We notice in Fig.~\ref{z_root_s} that $\zeta$ (a) as well as $\zeta/s$ (c) 
are decreasing with $\sqrt{s}$, which is qualitatively agreeing with the 
results of earlier studies~\cite{Kadam1,Kadam2}. 
The decreasing trend of $\zeta$ and $\zeta/s$ with $\sqrt{s}$ can 
be understood from the fact that $\mu_B$ decreases with $\sqrt{s}$ while $T$ remains 
fairly constant in the range of $\sqrt{s}$ analyzed here and according to
Fig.~\ref{z_s_mu}(a) and (c), the $\zeta$ and $\zeta/s$ decreases with 
decreasing of $\mu_B$.
%but at low $\sqrt{s}$, the solid line is only exhibiting a mild valley structure  around
%15-20 GeV. It indicates that the detailed $\tau(\vk,T,\mu_B)$ is responsible for this
%mild valley structure. The low $\sqrt{s}$ $(<20$ GeV) corresponds to 
%$\mu_B\sim 0.420$-$0.200$ GeV and $T\sim0.140$-$0.160$ GeV and if one can track
%the $\zeta$ in this $T-\mu_B$ region from 
%Figs.~\ref{z_s_Tmu}(a) and \ref{z_s_mu}(a), then a decreasing nature will be found.
%Similarly the increasing nature of $\zeta(\sqrt{s})$ at high $\sqrt{s}$ ($>20$ GeV)
%can also be tracked from the Figs.~\ref{z_s_Tmu}(a) and \ref{z_s_mu}(a) in the regions
%$T\sim0.160$-$0.165$ GeV and $\mu_B\sim0.200$-$0.001$ GeV.
%Next, Fig.~\ref{z_root_s}(c) shows that $\zeta/s$ for both cases are displaying mild valley
%structure around $\sqrt{s}\sim15$-$20$ GeV . The
%mild valley structure in $\zeta/s$ for the case of constant $\tau$ is solely appeared 
%due to rapid increment of entropy density at low $\sqrt{s}$. While 
%the valley structure of $\zeta(\sqrt{s})$ for our calculated 
%$\tau(\vk,T,\mu_B)$ is amplified in the plot of $\zeta/s$ vs $\sqrt{s}$
%because of this rapid increment of entropy density at low $\sqrt{s}$. 
%
\begin{figure}
\begin{center}
\includegraphics[scale=0.35]{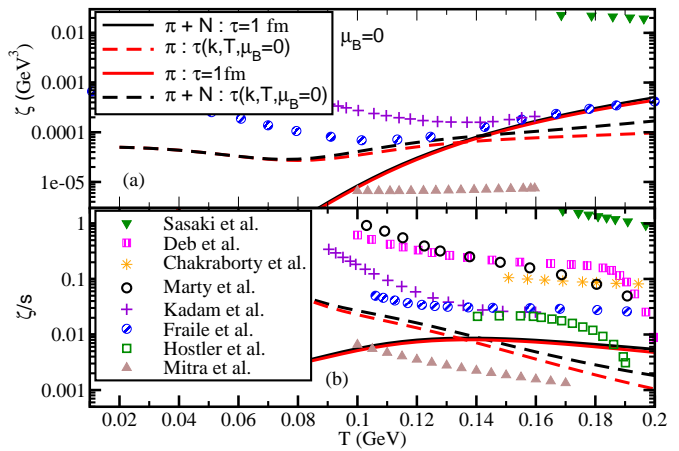}
\caption{(Color online) Our results of $\zeta$ (a) and $\zeta/s$ (b) vs $T$
at $\mu_B=0$ are compared with the earlier results of Sasaki et al. 
(Green triangles down~\cite{Sasaki}), Deb et al. (Pink solid squares~\cite{Kadam3}),
Chakraborty et al. (Brown stars~\cite{Purnendu}), Marty et al. (open circles~\cite{Cassing}),
Kadam et al. (Violet pluses~\cite{Kadam2}), Fraile et al. (Blue solid circles),
Hostler et al. (Open squares~\cite{Noronha})} 
\label{z_s_TComp}
\end{center}
\end{figure}

Fig.~\ref{z_s_TComp} is dedicated for comparative understanding of
our results with respect to the earlier investigations. As most of the 
works have been done at $\mu_B=0$, so we have plotted $\zeta$ (a) and 
$\zeta/s$ (b) against $T$ for $\mu_B=0$, where our results for $\pi$-
component (red lines) and $(\pi + N)$- components (black lines), using our
calculated $\tau(\vk, T, \mu_B=0)$ (dashed lines) and constant $\tau$ (solid lines),
are compared with the results, obtained by Sasaki et al. 
(Green triangles down~\cite{Sasaki}), Deb et al. (Pink solid squares~\cite{Kadam3}),
Chakraborty et al. (Brown stars~\cite{Purnendu}), Marty et al. (open circles~\cite{Cassing}),
Kadam et al. (Violet pluses~\cite{Kadam2}), Fraile et al. (Blue solid circles),
Hostler et al. (Open squares~\cite{Noronha}).
We see a large numerical band for $\zeta$ ($10^{-5}$-$10^{-2}$ GeV$^3$) or
$\zeta/s$ ($10^{-3}$-$10^{0}$), within which earlier estimations are located.
The results of the present work and Fraile et al.~\cite{Nicola} both show similar
kind of temperature dependence of $\zeta$ - it decreases at low $T$ 
domain ($<0.100$ GeV) and then increases at high $T$ domain ($>0.100$ GeV).
Monotonically increasing nature of $\zeta(T)$ for constant value of $\tau$ (solid lines)
discloses the fact that the origin of non-monotonic behavior of dashed lines are
because of explicit structure of $\tau(\vk,T,\mu_B=0)$.
The $\zeta(T)$ of Ref.~\cite{Kadam2} decreases up to $T\sim 0.150$
GeV after which a mild increment is observed. Most of the earlier 
works~\cite{Sasaki,Cassing,Kadam3,Dobado,Purnendu,Sarkar,Kadam2,Sarwar,Noronha,Nicola}
based on effective QCD model calculations~\cite{Sasaki,Cassing,Kadam3,Dobado,Purnendu}
as well as effective hadronic model calculations~\cite{Sarkar,Kadam2,Sarwar,Noronha,Nicola}
predicted a decreasing function of $\zeta/s (T)$ in the hadronic temperature domain, 
which is qualitatively similar with our results (dashed lines). 
These are not supporting the fact that $\zeta/s$ diverges or becomes large near
the transition temperature as indicated by Refs.~\cite{LQCD_zeta1,LQCD_zeta2,Tuchin},
within the temperature domain of quark phase. Some of the effective QCD model 
calculations~\cite{Defu,G_IFT,Kinkar,Dobado}, which can predict estimations of $\zeta/s$
in both temperature domain, exposed a peak structure near the transition temperature.
While some of the HRG model calculations~\cite{Kadam1,Noronha} have supported this behavior
by displaying an increasing tendency of $\zeta/s(T)$ as one goes towards the transition temperature
from the hadronic temperature domain. This kind of increasing $\zeta/s(T)$ is also
observed in our work when we consider the constant value of $\tau$ (solid lines).
Regarding this two opposite nature of $\zeta/s(T,\mu_B=0)$ within hadronic temperature domain,
Ref.~\cite{Noronha,Dobado} have exposed the possibility of both nature. Ref.~\cite{Noronha}
shows that inclusion Hagedorn states (HS) in HRG model can convert $\zeta/s(T)$ from decreasing
to increasing function. In this context, our results for explicit $T$, $\mu_B$ dependent $\tau$
and constant value of $\tau$ are also displaying both type of nature. Taking shear viscosity 
$\eta(T, \mu_B=0)$ from Ref.~\cite{G_eta_BJP}, based on same pion and nucleon
thermal fluctuations, we get $\zeta/\{(1/3-c_S^2)^2\eta\}\approx 5-4$ and 
$\zeta/\{(1/3-c_S^2)\eta\}\approx 0.8-0.7$. This is supporting the estimation
of gravity dual theory~\cite{Gravity} instead of the relation 
$\zeta/\{(1/3-c_S^2)^2\eta\}\approx 15$, followed by
photon fields~\cite{Weinberg}, scalar fields~\cite{scalar} or QCD theory~\cite{Arnold}.
So our estimation within the hadronic temperature domain is representing the strongly
coupled picture instead of weakly coupled scenario~\cite{Arnold}. Again, at high temperature
domain, our numerical values of $\zeta/s$ are matching (after extrapolation) 
with high temperature values of Refs.~\cite{Arnold,Kapusta2}- 
$\zeta/s(T\approx 0.200-0.400)\approx0.002-0.001$, obtained from the 
perturbative QCD calculations. 
In this regard, our estimation is indicating a smooth transformation from
the strongly coupled picture of the hadronic temperature domain to a weakly
coupled medium of quarks, instead of divergence or peak
structure of $\zeta/s$ near transition temperature.

%%%%%%%%%%%%%%%%%%%%%%%%%%%%%%%%%%%%%%%%%%%%%%%%%%%%%%%%%%%%%%%%%%%%%%%%%%
\section{Summary}
\label{sec:concl}
We have gone through a detailed microscopic calculation of
bulk viscosity coefficient for hadronic matter,
where thermodynamical equilibrium
conditions of all hadrons in medium have been treated by standard
HRG model, which is very successful to generate LQCD thermodynamics 
up to the transition temperature. 
%Owing to that success, HRG model
%can able to provide a detailed LQCD-like structure of trace anomaly
%in the non-perturbative domain of QCD, which is revealed by the 
%phase-space part of the bulk viscosity expression. 
The thermal widths
of medium constituents in the bulk viscosity expression inversely
determine their numerical strength. 
%
%Assuming pions and nucleons as
%most abundant medium constituents, we have concentrate on bulk viscosity
%calculations of pion and nucleon components, where their corresponding
%thermal widths are derived from their in-medium scattering probabilities
%with different mesonic and baryonic resonances in the hadronic matter.
%
Assuming pions and nucleons as most abundant medium 
constituents, we have concentrated on the
bulk viscosity contributions from pion and nucleon components,
where their corresponding thermal widths are derived
from their in-medium scattering probabilities with dif-
ferent mesonic and baryonic resonances in the hadronic matter.
Owing to the field theory version of optical theorem,
the imaginary part of pion and nucleon self-energy (on-shell) at finite temperature
give the estimation of their corresponding thermal widths.
In the one-loop diagrams of pion self-energy, we have taken different
mesonic and baryonic loops, while pion-baryon intermediate states
are considered in the one-loop diagrams of nucleon self-energy. 
%
%Their thermal widths are basically on-shell values of their corresponding
%Landau cut contributions, which are disappeared in absent of medium
%and therefore, these are inversely interpreted as their respective relaxation
%time, which proportionally control the numerical strength of transport coefficient
%like $\zeta$. 
%
Their thermal widths are basically
on-shell values of their corresponding Landau cut contributions, 
which disappear in the absence of medium and
therefore, these are inversely interpreted as their respective 
relaxation time, which proportionally control the numerical 
strength of transport coefficients like $\zeta$. 
Our result show that $\zeta(T)$ at $\mu_B=0$ increases in the 
high temperature domain ($0.080 <T (\rm{GeV})< 0.175$) but a decreasing
nature of $\zeta(T)$ has also been observed at low $T$ ($<0.08$ GeV). 
The $\pi\sigma$ and $\pi\rho$ loops of pion self-energy are respectively
responsible for the decreasing and increasing nature of $\zeta(T)$
at low and high $T$ domain. Addition of baryon loops in pion self-energy
mainly make $\zeta(T)$ reduce at high $T$ domain. Bulk viscosity
for nucleon component monotonically increases with $T$. At finite
$\mu_B$, the nucleon component of bulk viscosity is highly dominating over
the pion component. Adding nucleon and pion components, the total 
$\zeta$ increases with both $T$ and $\mu_B$. However, after dividing
by total entropy density, $\zeta/s$ appear as a decreasing function
of $T$ and with the variation of $\mu_B$, it increases first at 
low $\mu_B$ region and then decreases at high $\mu_B$ region. Along the 
beam energy axis, the $\zeta$ and $\zeta/s$ both decreases, as noticed in
some earlier works~\cite{Kadam1,Kadam2,Sarwar}.

%The $\zeta(\mu_B)$ as well as $\zeta/s(\mu_B)$ decreases 
%at low $\mu_B$ and then increases at high $\mu_B$ and hence a mild valley 
%structure is found in $\zeta$ vs $\mu_B$ and $\zeta/s$ vs $\mu_B$ plots. $\zeta/s(T)$ at $\mu_B=0$
%and $\mu_B\neq 0$ always exhibit decreasing nature. 
%$\zeta(\sqrt{s})$ and $\zeta/s(\sqrt{s})$ also exposes minima near 
%$\sqrt{s}\sim 15-20$ GeV. A rapid increment
%of entropy density has amplified the valley structure of $\zeta/s$
%near $\sqrt{s}\sim 15-20$ GeV, after which entropy density is generally
%saturated. 

During comparison with earlier results of $\zeta/s(T)$
at $\mu_B=0$, one can notice that the qualitative as well as quantitative 
nature is not a very settled issue. Some of them~\cite{LQCD_zeta1,LQCD_zeta2,Tuchin}
indicated divergence tendency of $\zeta/s$ near transition temperature,
some of effective QCD model calculations~\cite{Defu,G_IFT,Kinkar,Dobado} revealed
peak structure near transition temperature, whereas most of the 
effective QCD model calculations~\cite{Sasaki,Cassing,Kadam3,Dobado,Purnendu}
as well as effective hadronic model 
calculations~\cite{Sarkar,Kadam2,Sarwar,Noronha,Nicola}, including our present work,
predict a decreasing function of $\zeta/s (T)$ in the hadronic temperature domain,
with few exceptional HRG calculations~\cite{Kadam1,Noronha}.
Our decreasing $\zeta/s (T,\mu_B=0)$
is representing a strongly coupled picture in the hadronic temperature domain,
whose smooth extrapolation to high temperature domain agrees with a weakly
coupled picture~\cite{Arnold}.

%
%%%%%%%%%%%%%%%%%%%%%%%%%%%%%%%%%%%%%%%%%%%%%%%%%%%%%%%%%%%%%%%%%%%%%%%%%%%

{\bf Acknowledgment :}
During first and major part of this work,
SG is financially supported by the DST project with no NISER/R$\&$D-34/DST/PH1002,
(with title {\it ``Study of QCD phase Structure through high energy heavy ion collisions''}
and principal investigator Prof. B. Mohanty). During the last part of the work, 
SG is supported from UGC Dr. D. S. Kothari Post Doctoral Fellowship under
grant No. F.4-2/2006 (BSR)/PH/15-16/0060. SC acknowledges XIIth plan project no. 
12-R$\&D$-NIS-5.11-0300 and CNT project PIC XII-R$\&$D-VECC-5.02.0500 for support. 
SG thanks to high energy group of NISER
(Prof. B. Mohanty, Dr. A. Das, Dr. C. Jena, Dr. R. Singh, R. Haque, V. Bairathi, K. Nayak, V. Lyer, 
S. Kundu and others) and group of Calcutta University (Prof. A. Bhattacharyya, Prof. G. Gangopadhyay) 
for getting various academic and non-academic support at NISER and CU during 
this work and also to Dr. V. Roy and Prof. H. Mishra for some discussion regarding this work.
%
%
% 
%%%%%%%%%%%%%%%%%%%%%%%%%%%%%%%%%%%%%%%%%%%%%%%%%%%%%%%%%%%%%%%%%%%%%%%%%%%


\begin{thebibliography}{99}
%
\bibitem{KSS}P. Kovtun, D. T. Son, and O. A. Starinets, 
Phys. Rev. Lett. 94, 111601 (2005).
%
\bibitem{Tuchin} 
D. Kharzeev and K. Tuchin, 
J. High Energy Phys. 09 (2008) 093
%
\bibitem{Lat1}
A. Bazavov {\it et al}. (HotQCD Collaboration), 
Phys. Rev. D {\bf 90}, 094503 (2014). 
%
%\bibitem{Lat2}
%S. Borsanyi, Z. Fodor, S. D. Katz, S. Krieg, C. Ratti, and
%K Szab\'o, J. High Energy Phys. 01 (2012) 138.
%
\bibitem{LQCD_zeta1}H. B. Meyer, 
Phys. Rev. Lett. 100, 162001 (2008)
%
\bibitem{LQCD_zeta2} F. Karsch, D. Kharzeev, and K. Tuchin, 
Phys. Lett. B 663, 217 (2008).
%
%\bibitem{Heinz_rev} U. Heinz and R. Snellings, 
%Ann. Rev. Nucl. Part. Sci. 63, 123 (2013).
%
\bibitem{Torrieri} G. Torrieri and I. Mishustin, 
Phys. Rev. C 78, 021901 (2008).
%
\bibitem{Monnai} A. Monnai and T. Hirano, 
Phys. Rev. C 80, 054906 (2009).
%
\bibitem{Kodama} G. S. Denicol, T. Kodama, T. Koide and P. Mota, 
Phys. Rev. C 80, 064901 (2009).
%
\bibitem{Rajagopal} K. Rajagopal and N. Tripuraneni, 
JHEP 1003, 018 (2010).
%
\bibitem{Bozek} P. Bozek, 
Phys. Rev. C 81, 034909 (2010);
Phys. Rev. C 85, 034901 (2012);
P. Bozek and I. Wyskiel-Piekarska, 
Phys. Rev. C 85, 064915 (2012).
%
\bibitem{Heinz} H. Song and U. W. Heinz, 
Phys. Rev. C 81, 024905 (2010).
%
\bibitem{HM}J. Bhatt, H. Mishra, V. Sreekanth, 
J. High Energy Phys. 1011 (2010) 106;
Phys. Lett. B 704 (2011) 486;
Nucl. Phys. A 875 (2012) 181.
%
\bibitem{Dusling} K. Dusling and T. Schfer, 
Phys. Rev. C 85, 044909 (2012).
%
\bibitem{Victor}V. Roy and A.K.Chaudhuri,
Phys.Rev. C85 (2012) 024909.
%
\bibitem{Grassi1} J. Noronha-Hostler, G. S. Denicol, J. Noronha,
R. P. G. Andrade and F. Grassi, 
Phys. Rev. C 88, 044916 (2013).
%
\bibitem{Grassi2} J. Noronha-Hostler, J. Noronha and F. Grassi, 
Phys. Rev. C 90, no. 3, 034907 (2014).
%
\bibitem{Habich}M. Habich and P. Romatschke,
JHEP 12 (2014) 054.
%
\bibitem{Gale}S. Ryu, J.F. Paquet, C. Shen, G.S. Denicol, B. Schenke, S. Jeon, C. Gale,
Phys.Rev.Lett. 115 (2015) no.13, 132301
%
\bibitem{Arnold} P. Arnold, C. Dogan, G. D. Moore,
Phys.Rev. {\bf D 74}, 085021 (2006).
%
\bibitem{Sasaki}C. Sasaki and K. Redlich,
Phys. Rev. {\bf C 79}, 055207 (2009);
Nucl.Phys. {\bf A 832} (2010) 62.
%
\bibitem{Cassing}R. Marty, E. Bratkovskaya, W. Cassing, J. Aichelin, H. Berrehrah
Phys.Rev. C88 (2013) 045204.
%
\bibitem{Defu}S. Xiao, L. Zhang, P. Guo, D. Hou,
Chin. Phys. {\bf C 38} (2014) 054101.
%
\bibitem{G_IFT}S. Ghosh, T. C. Peixoto, V. Roy, F. E. Serna, G. Krein,
Phys. Rev. {\bf C 93} (2016) 045205.
%
\bibitem{Kinkar}K. Saha and S. Upadhaya, 
arXiv:1505.00177 [hep-ph].
%
\bibitem{Kadam3}P. Deb, G. Kadam, H. Mishra,
arXiv:1603.01952 [hep-ph].
%
\bibitem{Pratt}K. Paech and S. Pratt, 
Phys. Rev. C {\bf 74}, 014901 (2006).
%
\bibitem{Dobado}A. Dobado and J. Torres Rincon, 
Phys. Rev. D 86, 074021 (2012);
A. Dobado, F.J.Llanes-Estrada, J. Torres Rincon, 
Phys. Lett. B {\bf 702}, 43 (2011).
%
\bibitem{Purnendu} 
P. Chakraborty and J. I. Kapusta, 
Phys. Rev. C {\bf 83}, 014906 (2011).
%
\bibitem{Vinod} V. Chandra, 
Phys. Rev. {\bf D 84}, 094025 (2011); 
Phys. Rev. {\bf D 86}, 114008 (2012).
%  
\bibitem{Santosh}S. K. Das, J. Alam
Phys.Rev. D82 (2010) 051502.
%
\bibitem{Gavin}Gavin, S. 
Nucl.Phys. {\bf A 435} (1985) 826.
%
\bibitem{Sarkar}S. Mitra and S. Sarkar, 
Phys. Rev. {\bf D 87}, 094026 (2013);
S. Mitra, S. Gangopadhyaya, and S. Sarkar, 
Phys. Rev. {\bf D 91}, 094012 (2015)
%
\bibitem{Kadam1}G. P. Kadam, H. Mishra, 
Nucl. Phys. {\bf A 934} (2014) 133.
%
\bibitem{Kadam2}G. P. Kadam, H. Mishra, 
Phys. Rev. {\bf C 92} (2015) 035203;
Phys.Rev. C93 (2016) 025205.
%
\bibitem{Sarwar} G. Sarwar, S. Chatterjee, Jane Alam  
arXiv: 1512.06496[nucl-th].  
%
\bibitem{Noronha}
J. Noronha-Hostler, J. Noronha and C. Greiner, 
Phys. Rev. Lett. {\bf 103}, 172302 (2009).
%
\bibitem{Nicola} D. Fernandez-Fraile and A. Gomez Nicola,
Eur. Phys. J. C {\bf 62}, 37 (2009);
Phys. Rev. Lett. {\bf 102}, 121601 (2009).
%
\bibitem{pdg}K.A. Olive et al. (Particle Data Group), 
Chin. Phys. {\bf C, 38}, 090001 (2014). 
%
\bibitem{LQCDHRG1} A. Bazavov, T. Bhattacharya, M. Cheng, {\it et al.}, 
Phys. Rev. {\bf D 80}, 014504 (2009).
%
\bibitem{LQCDHRG2} S. Borsanyi {\it et al.},
J. High. Ener. Phys.   {\bf 1009}, 73  (2010).
%
\bibitem{LQCDHRG3} S. Borsanyi, Z. Fodor, S. D. Katz, S. Krieg,
C. Ratti, and K. Szabo, J. High. Ener. Phys. {\bf 1201}, 138 (2012).
%
\bibitem{GKS} S. Ghosh, G. Krein, S. Sarkar,
Phys. Rev. {\bf C 89} (2014) 045201.
%
%\bibitem{Zubarev}D. N. Zubarev
%{\it Non-equilibrium statistical thermodynamics}
%(New York, Consultants Bureau, 1974).
%
%\bibitem{Kubo} R. Kubo, 
%J. Phys. Soc. Jpn. {\bf 12}, 570 (1957).
%
%\bibitem{G_IJMPA} S. Ghosh,
%Int. J. Mod. Phys. {\bf A 29} (2014) 1450054.
%
\bibitem{Leopold} M. Post, S. Leupold, U. Mosel,
Nucl. Phys. {\bf A 741}, 81 (2004).
%
\bibitem{G_pi_JPG} S. Ghosh,
J. Phy. G 41, 095102 (2014).
%
\bibitem{G_eta_BJP}S. Ghosh,
Braz. J. Phys. 45 (2015) 687.
%
\bibitem{G_N} S. Ghosh,
%{\it The nucleon thermal width due to pion-baryon loops and its contributions in Shear viscosity}
Phys. Rev. {\bf C 90}, 025202 (2014).
%
\bibitem{G_NNst_BJP}S. Ghosh,
Braz. J. Phys. 44, 789 (2014).
%
\bibitem{HRGKarsch}F. Karsch and K. Redlich,
Phys. Lett. {\bf B 695}, 136-142 (2011)
%
\bibitem{Gravity}P. Benincasa, A. Buchel, and A. O. Starinets, 
Nucl. Phys. {\bf B 733}, 160 (2006); 
A. Buchel, 
Phys. Rev. D 72, 106002 (2005).
%
\bibitem{Weinberg}S. Weinberg, 
Astrophys. J. 168, 175 (1971).
%
\bibitem{scalar} 
R. Horsley and W. Schoenmaker, 
Nucl. Phys. {\bf B 280}, 716 (1987).
%
\bibitem{Kapusta2}J.~I.~Kapusta,
{\it Relativistic Nuclear Collisions}, Landolt-Bornstein New Series, Vol.\  I/23, 
ed.\  R.\ Stock (Springer-Verlag, Berlin Heidelberg 2010);
L. P. Csernai, J. I. Kapusta, and L. D. McLerran, 
Phys. Rev. Lett. {\bf 97}, 152303 (2006).

\end{thebibliography}
\end{document}